\documentclass{aa}
\usepackage{psfig}

\def\h2{H{\small II}}

\newcounter{qub}
\begin{document}

\title{Abundance patterns in the low-metallicity emission-line
galaxies from the Early Data Release of the Sloan Digital Sky Survey}

\author{Y. I. Izotov \inst{1}
\and G.\ Stasi\'nska \inst{2}
\and N. G.\ Guseva \inst{1}
\and T. X.\ Thuan\inst{3}}
\offprints{Y. I. Izotov, izotov@mao.kiev.ua}
\institute{      Main Astronomical Observatory,
                    Ukrainian National Academy of Sciences,
                    Zabolotnoho 27, Kyiv 03680,  Ukraine
\and
                    LUTH, Observatoire de Meudon, F-92195 Meudon Cedex, France
\and
                    Astronomy Department, University of Virginia,
                    Charlottesville, VA 22903, USA
}

\date{Received \hskip 2cm; Accepted}

\abstract{We have derived element abundances in 310 emission-line
galaxies from the Early Data Release of the Sloan Digital Sky Survey (SDSS) for
which the [O {\sc iii}]
$\lambda$4363 emission line was detected, allowing abundance determination by
direct methods. We found no extremely metal-deficient galaxy
($Z$ $\la$ $Z_\odot$/12)\thanks{12+log(O/H)$_\odot$=8.69 $\pm$ 0.05
(Lodders \cite{L03}).}, probably as a consequence of selection
effects in the SDSS sample. The oxygen abundance 12 + log O/H of the
SDSS galaxies sample lies in the range from $\sim$ 7.6 ($Z_\odot$/12) to
$\sim$ 8.4 ($Z_\odot$/2). This sample is merged with a sample of $\sim$100
blue compact dwarf galaxies with high quality spectra containing some
very low-metallicity objects to study the
abundance patterns of low-metallicity emission-line galaxies. We
find that the $\alpha$ element-to-oxygen abundance ratios do not show
any significant trends with the oxygen abundance, in agreement with previous
studies. The Fe/O abundance ratio is smaller than the solar value,
which we interpret as an indication that type Ia supernovae have not
yet appeared in these galaxies, implying an age of less than 1 -- 2
Gyr. However, a slight decrease of the Fe/O abundance ratio with 
increasing metallicity
suggests some depletion of iron onto dust in the galaxies with higher 
metallicities.
The N/O abundance ratio  ranges from log N/O= --1.6 to
--0.8. The fact that no galaxy with log N/O $\la$ --1.6 was
discovered implies that local low-metallicity emission-line galaxies
are of a different nature than  high-redshift damped Ly$\alpha$ systems with
log N/O of $\sim$ --2.3 and that their ages are probably larger
than 100 -- 300 Myr. Our data indicate  the existence of a gradual
nitrogen enrichment on a time-scale of a few Myr.
\keywords{galaxies: fundamental parameters --
galaxies: starburst -- galaxies: abundances}
}
\titlerunning{Abundance patterns in the emission-line
galaxies from the SDSS Early Data Release}


\maketitle


\section{Introduction}

The study of element abundances in  low-metallicity emission-line galaxies
is important for our understanding of the chemical evolution
of galaxies and for constraining models of stellar nucleosynthesis and the
shape of the initial mass function. The optical spectra of H {\sc ii}
regions in these galaxies show strong narrow emission lines superposed on a
stellar continuum that is rising toward the blue, allowing abundance
determinations of such heavy elements as nitrogen, oxygen, neon, sulfur,
argon, and iron.

The oxygen abundance  in the most metal-poor emission-line galaxies is around
1/30 times the solar value,
which makes these objects among the least
chemically evolved galaxies in the universe. It was proposed (Sargent \& Searle
\cite{SS70}) that the most metal-poor galaxies were actually
experiencing their first burst of star formation. However, subsequent
photometric and spectroscopic
studies have shown that the majority of metal-poor galaxies also possess
several Gyr old stellar populations. In the case of the two extremely
metal-deficient galaxies, I Zw 18 and SBS 0335--052, there is however 
no clear evidence of such stellar population and these galaxies may be 
genuinely young (e.g., Hunt, Thuan \& Izotov \cite{H03}; Guseva et al. 
\cite{G03c}).

Studies of local metal-poor emission-line galaxies are also crucial in the
context of galaxy formation. The proximity of these metal-poor galaxies allows
studies of
their structure, metal content, and stellar populations in a nearly pristine
environment with a sensitivity, precision, and spatial resolution that faint,
small angular size, distant high-redshift galaxies do not allow. In particular,
the comparison of the abundance patterns in the local low-metallicity dwarf
galaxies and distant damped Ly$\alpha$ systems  (DLA) can shed light on the
properties and the evolution of the primeval high-redshift galaxies.

The abundances of CNO  play an  especially
important role in the understanding the chemical evolution of the galaxies.
It is well established (e.g., Maeder \cite{M92}) that oxygen and likely
carbon are produced mainly by massive stars. The origin of nitrogen is more
controversial.

In local metal-poor
star-forming galaxies the nitrogen-to-oxygen abundance ratio  is found to be
constant (e.g., Garnett \cite{G90}) with a very low
dispersion around the mean value of log N/O = --1.6 for galaxies with oxygen
abundance 12 + log O/H $\la$ 7.6 (Thuan, Izotov \& Lipovetsky \cite{TIL95};
Izotov \& Thuan \cite{IT99}). This led Thuan et al. (\cite{TIL95}) and
Izotov \& Thuan (\cite{IT99}) to infer that in those galaxies,  nitrogen
has been produced by massive stars as a primary element and hence the 
lowest-metallicity galaxies
are young systems. However,  Henry, Edmunds \& K\"oppen (\cite{H00}) proposed
another explanation involving  a star formation occuring at a  very low rate,
with massive stars producing oxygen and intermediate-mass stars 
producing primary nitrogen with a typical time-lag of 250 Myr.

In high-redshift
low-metallicity damped Ly$\alpha$ systems, on the other hand, the spread of the
N/O abundance ratio appears to be very large (e.g., Lu,
Sargent \& Barlow \cite{L98}; Centuri\'on et al. \cite{C03}).
The N/O abundance ratio in some DLAs is found to be
one order of magnitude lower than that in blue compact dwarf galaxies  (BCD) of
same metallicity. If real, such differences
suggest that DLAs and BCDs are systems in different evolutionary stages.
This also raises doubts on the young age of the lowest-metallicity BCDs and
suggests that the production of nitrogen by intermediate-mass stars is actually
important in the low-metallicity BCDs
and DLAs.
This is in line with predictions of new
models of rotating stars (e.g., Meynet \& Maeder \cite{MM02}), although these
models also predict the production of a large amount of primary nitrogen by
low-metallicity massive stars.

Emission-line galaxies with very low metallicity are rare. Only two dozens
of BCDs with an oxygen abundance 12 + log O/H $\la$ 7.6 are known
up to now. The aim of this paper is to produce a large sample of
low-metallicity emission-line galaxies
with reliably derived element abundances and to compare them with those
in the high-redshift DLAs. This should allow a better understanding of the
nitrogen production and of the evolutionary status
of both types of objects. We have used the
early data release (EDR) of the Sloan Digital Sky Survey (SDSS)
(Stoughton et al. \cite{S02}) which includes the spectra of $\sim$ 50000
galaxies, and searched for low-metallicity emission-line galaxies. The
procedure and the resulting sample are described in section 2.  The derivation
of the elemental abundances is presented and discussed in section 3. The main
conclusions of this study are
presented in section 4.

\begin{figure}[t]
\hspace*{-0.0cm}\psfig{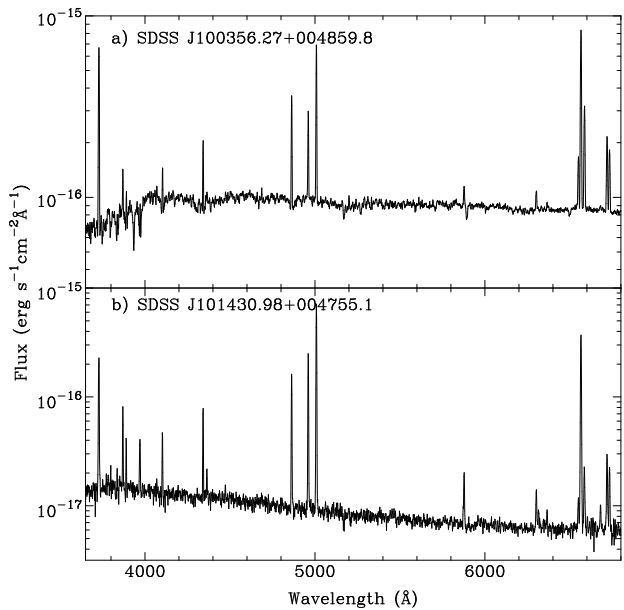}
\caption{Typical spectra of the (a) low-excitation and (b) high-excitation
H {\sc ii} regions from the SDSS EDR. The spectra are transformed to a
linear wavelength scale and zero redshift.}
\label{Fig1}
\end{figure}

\section{The sample}

The flux-calibrated spectra of the SDSS galaxies from the EDR have been
extracted
from the Space Telescope Science
Institute archives\footnote{http://archive.stsci.edu/sdss.}.

  From an examination of $\sim$ 50000 spectra, we selected
a sample of 310 emission-line galaxies with an emission feature at the
wavelength of the [O {\sc iii}] $\lambda$4363 emission line detected
at least at the 1$\sigma$ level. The presence
of this line allows a direct (as opposed to statistical or model-dependent)
determination of the electron temperature and abundances. It also discriminates
against metal-rich galaxies, in which important cooling of the H {\sc ii}
regions by metals strongly reduces the excitation of this line.
Because the wavelength scale of the extracted spectra is nonlinear they
have been transformed to a linear wavelength scale. Then spectra have been
reduced to a zero redshift using the galaxy
redshifts provided in the EDR. 
Typical spectra of the galaxies from the SDSS sample are shown in
Fig. \ref{Fig1}. The upper spectrum is of lower excitation
as compared to the lower spectrum as evidenced by the fluxes of [O {\sc ii}]
$\lambda$3727 and [O {\sc iii}] $\lambda$4959, 5007 emission lines relative
to H$\beta$.
Emission line fluxes have been measured using Gaussian fitting.
They have been corrected for  interstellar extinction
using the observed hydrogen Balmer line fluxes. The spectra in the
SDSS sample cover  the
wavelength range $\sim$ $\lambda$3820 -- 9300\AA, therefore the line
[O {\sc ii}] $\lambda$3727 has not been detected in low-redshift
galaxies with $z$ $\la$ 0.025. In those cases we used the
[O {\sc ii}] $\lambda$7320,7331 emission lines. Another problem with the
SDSS spectra is that in many cases some of the strongest lines, most often
H$\alpha$ and [O {\sc iii}] $\lambda$5007, are clipped. Therefore,
many objects with strong emission lines were not included in our sample.
However, in the cases when only H$\alpha$ and [O {\sc iii}] $\lambda$5007
emission lines are clipped, their fluxes are adopted to be respectively
2.8$\times$$I$(H$\beta$) and 3$\times$$I$([O {\sc{iii}}] $\lambda$4959).

\begin{figure}[t]
\hspace*{-0.0cm}\psfig{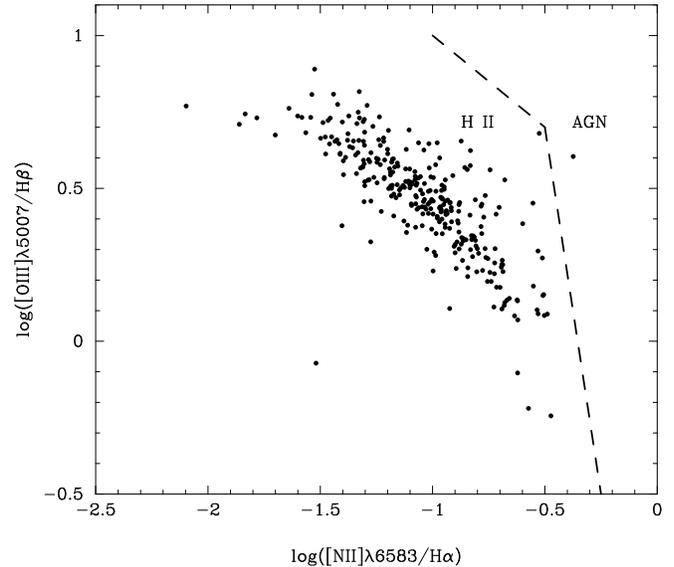}
\caption{Diagnostic diagram [O {\sc iii}]$\lambda$5007/H$\beta$ vs
[N {\sc ii}]$\lambda$6583/H$\alpha$ for low-metallicity emission-line galaxies 
from the EDR of the SDSS. The dashed line
separates the H {\sc ii} regions from galaxies ionized by a 
non-thermal radiation (labeled ``AGN'') (Osterbrock \cite{O89}).}
\label{Fig2}
\end{figure}

Figure \ref{Fig2} shows the classical diagnostic diagram [O {\sc
iii}] $\lambda$5007/H$\beta$ vs
[N {\sc ii}] $\lambda$6583/H$\alpha$ used to distinguish objects ionized by
massive main sequence stars from objects ionized by  non-thermal 
radiation.
The dashed line  separates the former (labeled ``H II'') from the latter 
(labeled ``AGN'') (Osterbrock \cite{O89}). Most of 
our selected objects  fall in the ``H II'' region on the diagram.
Only in two galaxies the ionization by a non-thermal radiation is important.
Note that the situation is very different when considering an entire sample of
emission-line galaxies without imposing the detection of [O {\sc iii}]
$\lambda$4363 emission line. For such a sample, which contains a large
proportion of metal-rich galaxies, about 30\% of the galaxies are found in the
AGN region (Heckman \& Kauffmann \cite{HK03}).

\begin{figure}[t]
\hspace*{-0.0cm}\psfig{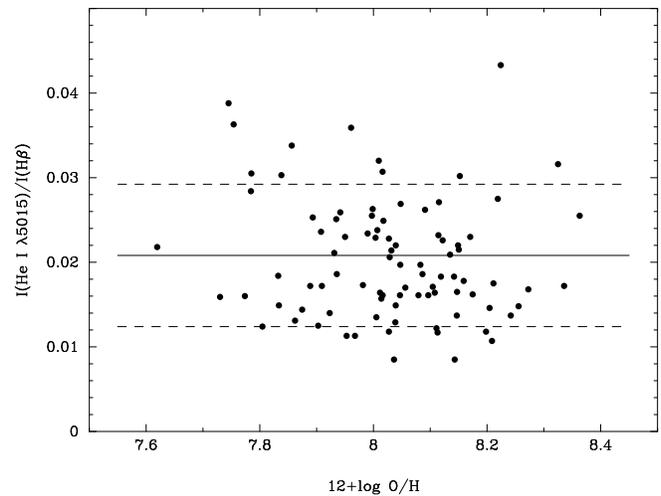}
\caption{He {\sc i} $\lambda$5015/H$\beta$ emission line flux ratio
as a function of the oxygen abundance. The solid line is the mean ratio and the
dashed lines are 1$\sigma$ alternatives.}
\label{Fig3}
\end{figure}

Because of the good spectral resolution of the SDSS spectra
the singlet He {\sc i} $\lambda$5015 emission line is separated from
the strong [O {\sc iii}] $\lambda$5007 emission line. This feature of the
SDSS spectra is of interest for the determination of the primordial He
abundance (Cota \& Ferland \cite{CF88}).
The flux of the He {\sc i} $\lambda$5015 line for the case B model is 
$\sim$ 0.02 that of H$\beta$,
while it is much weaker in the case A, only $\sim$ 0.0005 that of H$\beta$
(Brocklehurst \cite{B72}; Aller \cite{A84}). In Fig. \ref{Fig3} we show
the He {\sc i} $\lambda$5015/H$\beta$ emission line flux ratio vs oxygen
abundance 12 + log O/H. The solid line is the mean value of the
He {\sc i} $\lambda$5015/H$\beta$ emission line flux ratio, and dashed
lines show 1$\sigma$ deviations.
The points in Fig. \ref{Fig3} are scattered around a mean
value of $\sim$ 0.02, consistent with the case B model for He in singlet
states.

In addition to the SDSS EDR, we use a sample of galaxies 
which has been collected primarily to study the helium abundances in
low-metallicity blue compact dwarf galaxies (the HeBCD sample). This sample 
is gradually increasing with the addition
of new objects and now includes $\sim$ 100 galaxies. Most of the objects are 
blue compact dwarf galaxies from the First and Second Byurakan surveys, but a 
number of them are emission-line galaxies from the Hamburg, University of 
Michigan, Tololo and Case
surveys. High signal-to-noise ratio spectra of these galaxies in the
wavelength range $\lambda$3600 -- 7400\AA\ have been obtained with different
2m - 10m class telescopes. Some galaxies were observed several times.
We included in the sample all these independent observations.
All spectra were reduced in the same way according to the prescriptions of 
Izotov et al. (\cite{ITL94,ITL97}). The line fluxes corrected for extinction 
can be found in Izotov et al. (\cite{ITL94,I96,ITL97,I97,I99,I01a,I01b}),
Izotov \& Thuan (\cite{IT98a,IT98,IT03}), Thuan et al. (\cite{TIL95,TIF99}),
Lipovetsky et al. (\cite{Li99}), Guseva et al. (\cite{G00,G01,G03a,G03b,G03c}),
Hopp et al. (\cite{Ho00}), Fricke et al. (\cite{F01}).

The comparison of the global characteristics of the SDSS and HeBCD samples
was presented by Stasi\'nska \& Izotov (\cite{SI03}) and can be
summarized as follows. The SDSS galaxies are generally fainter than the HeBCD 
ones and their spectra are of lower signal-to-noise ratio.
On average, the SDSS galaxies are more distant but their
H$\beta$ luminosities are similar to those of the HeBCD galaxies. The
equivalent width distributions are different, being more skewed
toward large equivalent widths in the HeBCD sample.
Indeed, many galaxies from the HeBCD sample have been observed with the aim
of deriving the primordial helium abundance, thus objects with
large equivalent widths of the emission lines were preferentially selected. The
distributions of
$E(B-V)$ are roughly similar in both samples. For the majority of the objects,
the extinction is small ($E(B-V)\la0.2$).

\section{Element abundance determination}

\begin{figure*}[t]
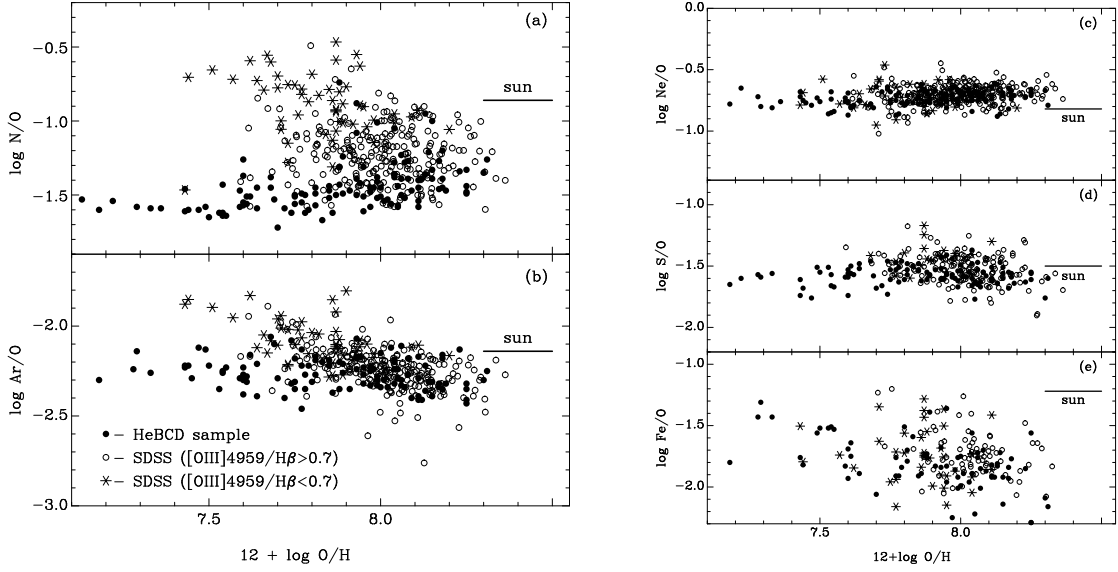

\hspace*{-0.0cm}\psfig{figure=0101.f4a.ps,angle=0,height=7.5cm,clip=}
\hspace*{1.0cm}\psfig{figure=0101.f4b.ps,angle=0,height=7.5cm,clip=}
\caption{log N/O (a), log Ar/O (b), log Ne/O (c), log S/O (d) and log Fe/O (e)
vs oxygen abundance
12 + log O/H for the HeBCD and SDSS galaxies. Filled circles are the
HeBCD galaxies, open circles are the SDSS galaxies with [O {\sc iii}]
$\lambda$4959/H$\beta$ $\geq$ 0.7 and asterisks are the SDSS galaxies with
[O {\sc iii}] $\lambda$4959/H$\beta$ $<$ 0.7. Solar abundance ratios from
Lodders (\cite{L03}) are shown by solid lines.}
\label{Fig4}
\end{figure*}

\begin{figure*}[t]
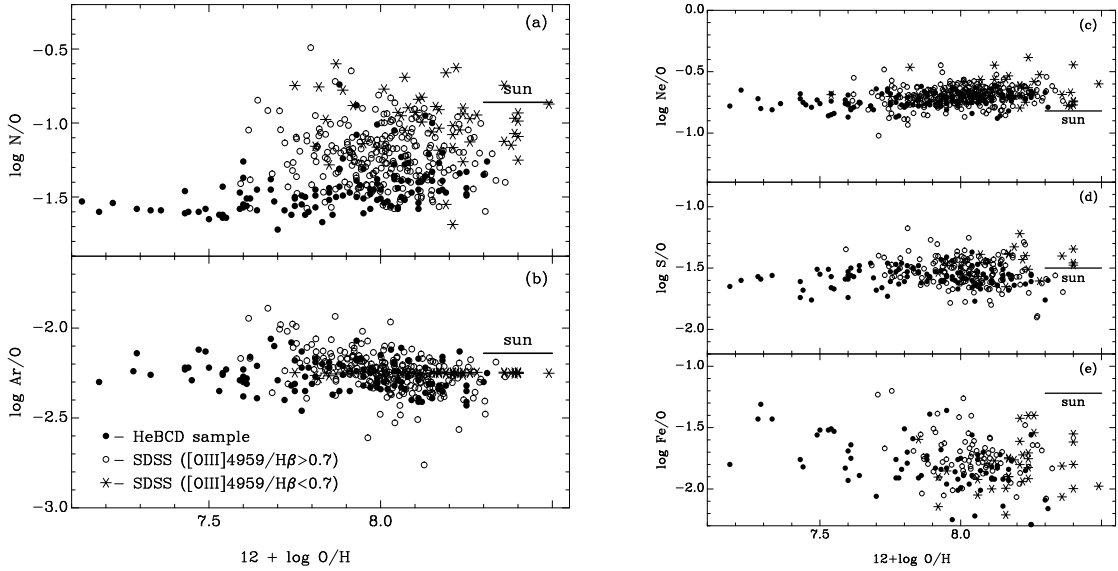

\hspace*{-0.0cm}\psfig{figure=0101.f5a.ps,angle=0,height=7.5cm,clip=}
\hspace*{1.0cm}\psfig{figure=0101.f5b.ps,angle=0,height=7.5cm,clip=}
\caption{Same as in Fig. \ref{Fig4}, but after correction of the electron
temperature $T_e$(O {\sc iii}) in the SDSS galaxies with [O {\sc iii}]
$\lambda$4959/H$\beta$ $<$ 0.7 (asterisks).
}
\label{Fig5}
\end{figure*}

The physical conditions and element abundances are derived from the emission
line fluxes using the same prescriptions as in Izotov et al.
(\cite{ITL94,ITL97}). The electron temperature $T_e$(O {\sc iii}) and
number density $N_e$(S {\sc ii}) are derived
from the [O {\sc iii}] $\lambda$4363/($\lambda$4959+$\lambda$5007) and
[S {\sc ii}] $\lambda$6717/$\lambda$6731 emission line flux ratios.
The quality of the spectra is good enough to derive the
N, O, Ne, S, Ar and Fe abundances. We use [N {\sc ii}] $\lambda$6548, 6583
emission lines for the determination of nitrogen abundance, [O {\sc ii}] 
$\lambda$3727 or [O {\sc ii}] $\lambda$7320,7331,
[O {\sc iii}] $\lambda$4959, 5007 for the oxygen abundance, [Ne {\sc iii}]
$\lambda$3868 for the neon abundance, [S {\sc iii}] $\lambda$6312 and  [S {\sc
ii}] $\lambda$6717, 6731 for the
sulfur abundance when all these lines are detected, [Ar {\sc iii}]
$\lambda$7135 (and [Ar {\sc iv}] $\lambda$4740 if seen) for the argon
abundance, [Fe {\sc iii}] $\lambda$4658, 4988 for the iron
abundance. The ionization correction factors to obtain the elemental abundances
from the ionic ones are the same as  used by Izotov et al.
(\cite{ITL94,ITL97}). We have checked the compatibility of these ionization
correction factors with those given by the  photoionization models of
Stasi\'nska \& Izotov (\cite{SI03}) which were built to fit the
observed spectral
properties of the samples analyzed in the present paper. We have also run
additional sequences of photoionization models, taking into account
dielectronic
recombination for sulfur and argon ions, and using Starburst 99 (Leitherer et
al. \cite{L99}) with the stellar model atmospheres described in
Smith et al. (\cite{S02}). 
We find that the ionization correction factors that have been used are in
reasonable agreement with the predictions from the models. The only noteworthy
difference is in the case of neon. To compute the abundances, it was assumed
that Ne/O=Ne$^{++}$/O$^{++}$ (which is the usual assumption). However, our
models show that, as the ionization parameter decreases (or, equivalently, as
the O$^{++}$/O$^{+}$ ratio decreases), the proportion of Ne$^{++}$ with respect
to O$^{++}$ becomes larger, since charge transfer between O$^{++}$ and
H$^{0}$ becomes more efficient to recombine O$^{++}$. This effect is noticeable
only when O$^{++}$/O$^{+}$ is smaller than 8, and can lead
to an overestimate of Ne/O by 0.2 dex when O$^{++}$/O$^{+}$ is equal to 1.
However, the
importance of this effect also depends on the spectral energy distribution of
the ionizing radiation field, which, in turn, depends on the metallicity. We
have thus decided, for simplicity, to maintain the usual way to derive the Ne/O
ratio but keeping in mind any possible bias for discussion.

In Fig. \ref{Fig4} we show the computed N/O, Ar/O, Ne/O, S/O and Fe/O abundance
ratios
vs oxygen abundance 12 + log O/H for the  SDSS and HeBCD galaxies. The
HeBCD galaxies are represented by filled circles, while the SDSS
galaxies are
divided into galaxies with [O {\sc iii}] $\lambda$4959/H$\beta$ $\geq$ 0.7
(represented by open circles) and galaxies with [O {\sc iii}]
$\lambda$4959/H$\beta$ $<$ 0.7 (asterisks). In the latter galaxies, the
equivalent width of  H$\beta$ is generally low and the [O {\sc iii}]
$\lambda$4363
emission line weak. This may cause a substantial uncertainty
in the placement of the continuum due to the absorption lines
of H$\gamma$, He {\sc i} and some heavy elements blueward and redward of the
[O {\sc iii}] $\lambda$4363 line. Consequently, we suspect that the derived
abundances may be biased.
As seen in Fig. \ref{Fig4}a, among the SDSS galaxies with [O {\sc iii}]
$\lambda$4959/H$\beta$ $\geq$ 0.7 no object was found to have 12 +  log O/H
$\la$ 7.6. The only SDSS galaxies with such a low oxygen abundance have
[O {\sc iii}] $\lambda$4959/H$\beta$ $\la$ 0.7 (labeled by asterisks). 
Note that all the SDSS galaxies for which 12 + log
O/H $\la$ 7.6  have N/O values which are several times larger than the N/O
ratios derived in the most metal-deficient galaxies from the HeBCD sample.
Furthermore, the asterisks show a clear
trend of Ar/O decreasing with increasing oxygen abundance (see Fig.
\ref{Fig4}b), whereas the Ar/O abundance ratio is expected to be constant
over the whole range of oxygen abundance considered here 
(Izotov \& Thuan \cite{IT99}).

This suggests that
the derived oxygen abundances are underestimated, possibly due to an
overestimation of the very weak [O {\sc iii}] $\lambda$4363 line whose
flux in these galaxies does not exceed 2\% -- 4\% of the H$\beta$ flux.
The effect of an error in the electron temperature is much weaker on the
determination of the Ne/O and S/O abundance ratios than on the determination of
the N/O and Ar/O ones. This explains why no trend is seen in the
log Ne/O vs 12 + log O/H or log S/O vs 12 + log O/H  diagrams
(from nucleosynthesis, Ne/O and
S/O are expected to be constant, similarly to Ar/O). For example, if the real
electron temperature in a galaxy is 10000\,K while the
[O {\sc iii}] $\lambda$4363/$\lambda$4959 ratio
gives an electron temperature of 20000\,K, the N/O ratio will be 
overestimated by
a factor of about 3 if derived with [O {\sc ii}] $\lambda$3727 and about 6 if
derived with [O {\sc ii}] $\lambda$7330, the Ar/O ratio will be
overestimated by a factor of about 2, while Ne/O and S/O will be
underestimated by only 30\%. Note that such an important overestimation of the
[O {\sc iii}] $\lambda$4363 flux is quite likely
in noisy or low excitation spectra, due to a bad placement of the continuum or
due to an erroneous detection of the [O {\sc iii}] $\lambda$4363 feature.
However, we note that the auroral [S {\sc iii}] $\lambda$6312 emission line
has been detected in 30\% of the spectra with [O {\sc iii}]
$\lambda$4959/H$\beta$ $<$ 0.7 implying that the presence of the
[O {\sc iii}] $\lambda$4363 emission line is not an artefact in those spectra.
It is likely that some additional heating mechanism other than the
stellar radiation is important in the galaxies labeled by asterisks.

\begin{figure*}[t]
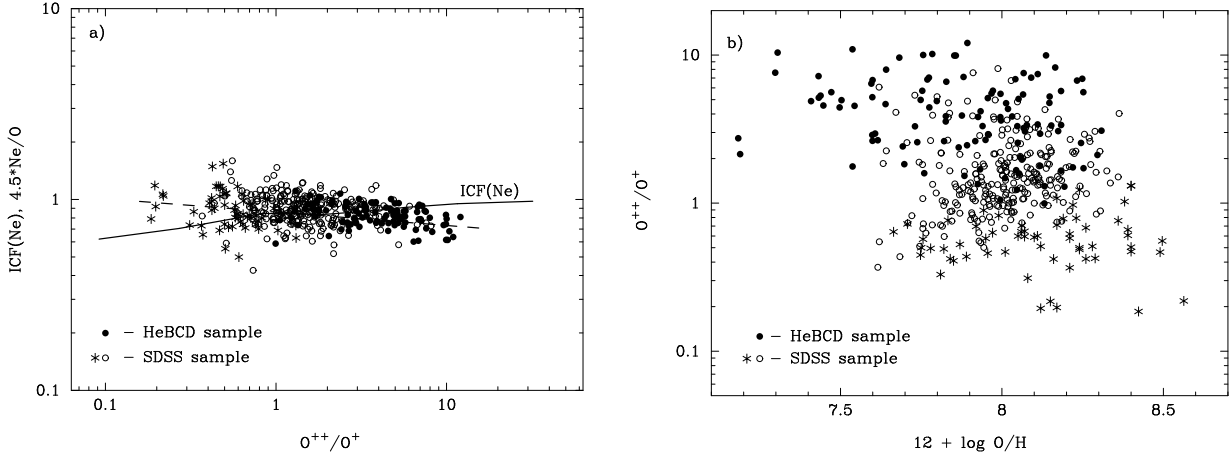

\hspace*{-0.0cm}\psfig{figure=0101.f6a.ps,angle=270,height=6.cm,clip=}
\hspace*{0.5cm}\psfig{figure=0101.f6b.ps,angle=270,height=5.9cm,clip=}
\caption{
(a) The Ne/O abundance ratios for our sample galaxies as a function of
O$^{++}$/O$^{+}$ ratio (same symbols as Fig. \ref{Fig4}). The dashed line shows
the trend of the data derived from the observations. The solid line
shows the ionization correction factor $ICF$(Ne) as derived from 
photoionization models with
metallicity $Z$ = $Z_\odot$/5 (see text). (b) The observed values of
O$^{++}$/O$^{+}$ as a function of 12 + log O/H.}
\label{Fig6}
\end{figure*}

To correct the bias in the abundance determinations due to this temperature
effect, we assume that all the SDSS galaxies labeled with asterisks in Fig.
\ref{Fig4} have an Ar/O abundance ratio equal to the mean value 
(log Ar/O = --2.25) derived
by Izotov \& Thuan (\cite{IT99}) for the galaxies from the HeBCD sample.
Then we compute the value of the electron
temperature needed to achieve such a result,  and use this 
temperature to rederive the abundances of O, N, Ne, S and Fe.  The 
results of
this procedure are shown in
Fig. \ref{Fig5}, which shows the same diagrams as Fig. \ref{Fig4} but with the
corrected abundances. We see that, now, there is not a single SDSS galaxy with
12 + log O/H $\la$ 7.6. Also, the location of the entire sample of
SDSS galaxies in the log N/O vs
12 + log O/H diagram is substantially more consistent with the distribution
of the galaxies from the HeBCD sample (Fig. \ref{Fig5}a), since no galaxies
with
12 + log O/H $<$ 7.6 and high N/O abundance ratio are found. On the other hand,
no significant changes appear in the Ne/O, S/O and Fe/O vs 12 + log O/H 
diagrams except that the galaxies represented by asterisks are shifted towards 
higher oxygen abundance with respect to Fig. \ref{Fig4}. This supports the 
correctness of our approach.

\section{Discussion}

The distribution of the SDSS galaxies from the Early Data Release (open circles
and asterisks) in the different panels of Fig. \ref{Fig5} confirms previous
findings for HeBCD galaxies (Thuan et al. \cite{TIL95};
Izotov \& Thuan \cite{IT99}). The only striking difference is that no extremely
metal-deficient galaxies with 12 + log O/H $<$ 7.6 ($Z$ $<$ $Z_\odot$/12) are
found in the SDSS sample. Recall however that we have excluded from
consideration
galaxies with strong but clipped emission lines. Some of those galaxies
could actually be metal-poor. Meanwhile, the SDSS Data Release 1 (DR1)
has become available\footnote{http://www.sdss.org/dr1.}.
Galaxies from the EDR have re-reduced spectra included in the DR1.
Among galaxies from the DR1 sample only a few have spectra with clipped
emission lines. Therefore, using the data from the DR1 sample, we have checked
whether very metal-poor galaxies are present in the EDR, among those
which have been
excluded from our consideration because of clipped lines. No
such galaxies were found. We suspect that the  selection criteria
used in the SDSS to extract galaxies from multi-colour imaging data
for follow-up spectroscopic work
(Stoughton et al. \cite{S02}) are biased against the most metal-poor
galaxies. First, in the SDSS galaxies are distinguished from
stars by morphology. But distant blue compact dwarf galaxies can be
morphologically indistinguishable from stars, and some of them might even
appear in the quasar sample of the SDSS. In fact, we find several
cases when low-metallicity emission-line galaxies were classified as
QSOs.
Second, extremely metal-poor galaxies are expected to be of very low luminosity
because of the metallicity-luminosity relation (e.g.,
Skillman, C\^ot\'e \& Miller \cite{S03}) and for that reason they are rejected 
from the list of targets for SDSS spectroscopy. The SDSS aims at obtaining a
complete sample of galaxies brighter than a Petrosian magnitude $r_*$ = 17.77
(see Stoughton et al. \cite{S02}).
Galaxies with an absolute brightness of $\sim$ --14 mag, similar to 
that of I Zw 18, are already fainter than the completeness limit at a 
redshift of 0.01.
The selection criteria for the HeBCD galaxies were completely 
different. The objects were chosen
from objective prism surveys to have large equivalent widths
of the H$\beta$ and [O {\sc iii}] $\lambda$4959, 5007 emission lines, 
mainly for the determination of the pregalactic helium abundance. 
Therefore the metallicity distributions of the two samples are expected 
to be different.


\begin{figure}[t]
\hspace*{-0.0cm}\psfig{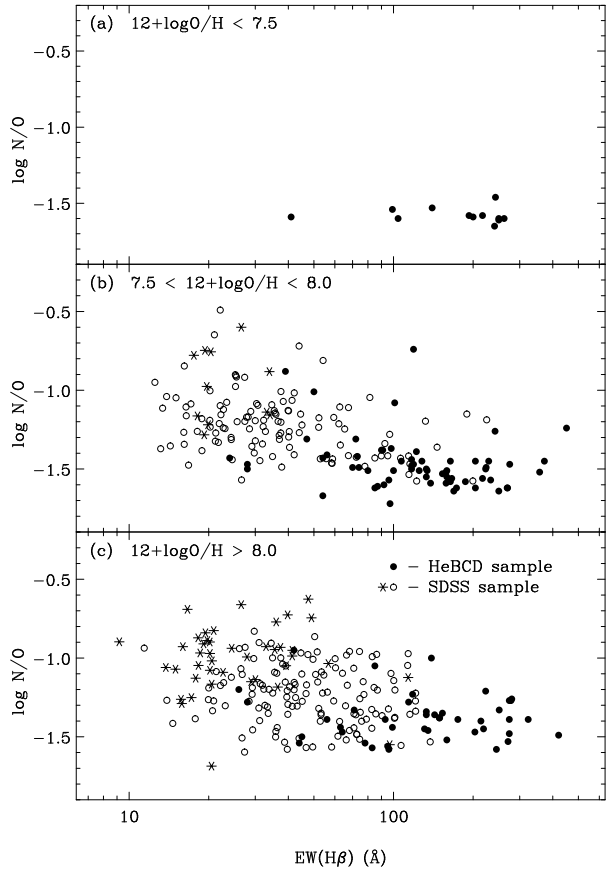}
\caption{Nitrogen-to-oxygen abundance ratio, log N/O, vs equivalent width
of H$\beta$ emission line, EW(H$\beta$), for the galaxies in three
metallicity bins defined as in Stasi\'nska \& Izotov (\cite{SI03}).
}
\label{Fig7}
\end{figure}

We find no significant trends with the oxygen abundance for
the Ne/O and S/O abundance ratios (Figs. \ref{Fig5}c -- \ref{Fig5}d),
although higher-metallicity galaxies have log Ne/O by $\sim$0.1 dex higher than
low-metallicity galaxies. This small trend is likely caused by
a bias in the Ne/O abundance determination, as discussed before. Formally, 
one can write
Ne/O = $ICF$(Ne)$\times$(Ne$^{++}$/O$^{++}$). As explained in Sect. 3, we have
adopted
$ICF$(Ne) = 1 whereas it is smaller than unity and decreases with
decreasing of O$^{++}$/O$^{+}$. In Fig. \ref{Fig6}a, we show with the
solid curve the value of $ICF$(Ne) as a function of O$^{++}$/O$^{+}$
for a sequence of photoionization models with metallicity $Z$ =
$Z_\odot$/5, computed using the stellar atmospheres from Smith et al.
(\cite{S02}), but otherwise identical to the sequence of models that matches
the ``high metallicity bin'' of our sample galaxies (cf.
Stasi\'nska \& Izotov \cite{SI03} for more details). We see that
indeed $ICF$(Ne) is slightly smaller than unity at low excitation.
The effect on $ICF$(Ne) is even larger for lower metallicity, because
the stellar energy distribution in the Lyman continuum is then
harder. However, galaxies with lower metallicities tend to have
higher O$^{++}$/O$^{+}$, as seen in Fig. \ref{Fig6}b, therefore the
bias in the derived Ne/O ratio ends up being unimportant. In Fig.
\ref{Fig6}a, we have also plotted the derived Ne/O values for our
sample galaxies, and drawn as a dashed line the observed trend, which
is clearly opposite to the trend in $ICF$(Ne). The correction for $ICF$(Ne)
shown in Fig. \ref{Fig6}a would thus reduce this trend.

We find that
  the Fe/O abundance ratios are generally significantly smaller than 
solar and  similar to the ones derived
in the most metal-poor stars in the Galaxy (Caretta, Gratton \& Sneden
\cite{C00}). This may argue in
favour either of a moderate depletion of iron onto dust grains or of
production of iron being dominated by  type II supernovae (arising
from massive stars). However we note a slight decrease of the Fe/O 
abundance ratio with increasing metallicity (by $\sim$ 0.2 dex when 
the oxygen abundance increases by one order of
magnitude as seen in Fig. \ref{Fig5}e). This implies that some 
depletion occurs in
the high-metallicity galaxies. Nevertheless, the similarity of the 
Fe/O abundance
ratios in the studied galaxies and halo stars favours
the production of iron being dominated by  type II supernovae, which implies
an age of $<$ 1 -- 2 Gyr, i.e. before the contribution of type Ia supernovae
(which arise from intermediate mass stars) becomes important (Thuan et al.
\cite{TIL95}; Izotov \& Thuan \cite{IT99}).

We now compare the abundance patterns we find for the local metal-poor
emission-line galaxies
with those found for DLAs
(e.g., Centuri\'on et al. \cite{C03}). Many DLAs actually
follow the N/O vs O/H distribution of the local dwarf emission-line galaxies.
This suggests
similar star formation histories and evolutionary status for these two classes
of objects.
However, we have found no local counterparts of the high-redshift
DLAs with log N/O
$\la$ --1.6. It has been proposed (e.g., Centuri\'on et al. \cite{C03})
that  DLAs with low N/O abundance ratio are young systems just starting
to form stars and where both N and O are produced by massive stars, while
DLAs with high N/O abundance ratio as well as local emission-line galaxies are
older, with an age of
at least several hundred Myr, to allow the intermediate-mass stars
to deliver their nitrogen. This
conclusion is in line with predictions of the evolution of
rotating stars (Meynet \& Maeder \cite{MM02}). However, such an interpretation
presents problems. In the N/O vs O/H plane, DLAs with high N/O and DLAs with
low N/O
abundance ratios form two discrete groups. It is not clear, as
discussed by Centuri\'on et al. (\cite{C03}), why only very young
DLAs with low N/O  and relatively evolved DLAs with high N/O  are found,
while no intermediate case is known. New observations and studies of a larger
sample of  DLAs and local emission-line galaxies are necessary to clarify
the situation.

If the nitrogen and oxygen production scenario discussed above
holds, then the low N/O dispersion in the galaxies
with low oxygen abundances in
Fig. \ref{Fig5}a can be explained by massive stars continuously producing 
oxygen and intermediate-mass stars continuously producing primary nitrogen. 
However, the large spread of the N/O abundance ratios
at higher metallicities needs to be explained. In Fig. \ref{Fig7} we show
the N/O abundance ratios as a function of the
H$\beta$ equivalent width EW(H$\beta$). The samples are split into
three metallicity bins following Stasi\'nska \& Izotov (\cite{SI03}).
No trend is seen in the low-metallicity bin (but the number of the
galaxies and the range of EW(H$\beta$) are small). On the other hand, clear
trends are present for the galaxies in the intermediate- and high-metallicity 
bins for which the N/O abundance ratio and its dispersion increase when
EW(H$\beta$) decreases.
Since EW(H$\beta$) is an indicator of the age of the latest burst of star
formation in the galaxy (but see the discussion by Stasi\'nska \&
Izotov \cite{SI03}), the trends and large dispersions of
N/O in the intermediate- and high-metallicity bins (Fig. \ref{Fig7}) can
be explained by the additional production of nitrogen in massive Wolf-Rayet
stars arising from the latest burst,  on top of the nitrogen
produced by intermediate-mass stars. This effect is expected to be
lower for the H {\sc ii} regions in the low-metallicity bin because of
a significantly lower fraction of Wolf-Rayet stars at low
metallicities. Note, however, that a few cases of extremely
low-metallicity galaxies have been reported in the literature with
log N/O slightly larger than --1.6 (van Zee \cite{V00}; 
Skillman et al. \cite{S03}).
These objects have EW(H$\beta$) of the order of 100\AA, therefore the most
massive stars from the latest burst have probably had time to evolve and
enrich the nebula in nitrogen. A larger sample of extremely
low-metallicity galaxies would be welcome to study this issue in more
detail.

\section{Conclusions}

We have discussed in this study the abundances of the heavy elements in the
310 emission-line galaxies from the sample of the Sloan Digital Sky Survey
(SDSS) Early Data Release (EDR)
with a detected [O {\sc iii}] $\lambda$4363 emission line, allowing a direct
element abundance determination based on the electron temperature.
We have shown that the abundances are biased for galaxies with spectra of low
excitation, and we attribute this bias mainly to a false detection of
[O {\sc iii}] $\lambda$4363 or to a wrong placement of the continuum due to a low
signal-to-noise ratio and adjacent stellar absorption lines. However, we cannot
exclude that an additional heating mechanism different from the stellar
radiation may be important for low-excitation galaxies.
We have corrected for this bias
by requiring that the Ar/O ratio be equal to the expected value.
After correcting for this bias and merging the SDSS sample with the sample
of the galaxies with large equivalent widths of emission lines (HeBCD sample),
we obtain a sample of $\sim$ 400
emission-line galaxies in total. Our main conclusions are as follows:

1. Despite an examination of the entire SDSS EDR sample of galaxies,
we found no
galaxy with extremely low metallicity (12 + log O/H $<$ 7.6, i.e. $Z$ $<$
$Z_\odot$/12) while the HeBCD sample 
which consists of $\sim$100 galaxies
contains about 15 such galaxies. We suspect that the spectroscopic
data base of the SDSS actually discriminates against the most
metal-poor galaxies.

2. Emission-line galaxies from the SDSS sample in general follow the
distributions of the N/O, Ne/O, S/O, Ar/O, Fe/O abundance ratios vs oxygen
abundance found earlier for the galaxies in the HeBCD sample (Thuan et al.
\cite{TIL95}; Izotov \& Thuan \cite{IT99}).

3. The $\alpha$ element-to-oxygen abundance ratios do not show any
significant trend with oxygen abundance.

4. The Fe/O abundance ratio shows a significant underabundance of iron
relative to oxygen as compared to solar, suggesting  either
depletion of iron onto dust grains or a dominant production
of iron in the massive supernovae of the type II. We favour the
latter explanation, which implies an age of less than $\sim$ 1 -- 2 Gyr.

5. No galaxy with log N/O $\la$ --1.6 was found suggesting that
local low-metallicity emission-line galaxies are different from those
as  high-redshift DLAs with the considerably lower
log N/O of $\sim$ --2.3. These DLAs are considered to be
truly young galaxies, with nitrogen produced only by massive stars. If this
interpretation is correct, then our sample of $\sim$ 400
dwarf emission-line galaxies contains no extremely young galaxy (i.e.
with an age
$\la$ 100 -- 300 Myr).

6. Our data indicate a gradual enrichment of the
galaxies in nitrogen by massive stars from the most recent starburst.
The effect is best seen among galaxies from our sample that have
intermediate metallicities.

\begin{acknowledgements}
Y. I. I. acknowledges the support of the University of Paris 7 and the
Observatoire de Paris, where part of
this work was carried out.
Y. I. I. and N. G. G. acknowledge the support of the Swiss SCOPE 7UKPJ62178 grant.
T. X. T. and Y. I. I. acknowledge the partial financial support of NSF
grant AST 02-05785.
All the authors acknowledge the work of the Sloan Digital Sky
Survey (SDSS) team. The SDSS is a joint project of The University of
Chicago, Fermilab, the Institute for Advanced Study, the Japan Participation
Group, the Johns Hopkins University, the Los Alamos National Laboratory, the
Max-Planck-Institute for Astronomy (MPIA), the Max-Planck-Institute for
Astrophysics (MPA), New Mexico State University, Princeton University, the
United States Naval Observatory, and the University of Washington.
Funding for the project has been provided by the Alfred P. Sloan Foundation,
the Participating Institutions, the National Aeronautics and Space
Administration, the National Science Foundation, the U.S. Department of Energy,
the Japanese Monbukagakusho, and the Max Planck Society.
\end{acknowledgements}

\end{document}